\documentclass[reprint,aps,prl,amsmath,amssymb,superscriptaddress,nofootinbib]{revtex4-2}

\usepackage{graphicx}
\usepackage{hyperref}
\usepackage{xcolor}
\usepackage{tikz}
\usepackage{pgfplots}
\usepackage{mathrsfs}
\pgfplotsset{compat=1.18}
\usetikzlibrary{arrows.meta,decorations.markings,decorations.pathmorphing}

\begin{document}

\title{The simplest event shape in $\mathcal{N}=4$ super Yang--Mills}

\author{Carlos Bercini}
\affiliation{Department of Mathematics, King's College London, Strand, London WC2R 2LS, United Kingdom}

\author{Sidhaarth Kumar}
\affiliation{Department of Mathematics, King's College London, Strand, London WC2R 2LS, United Kingdom}

\begin{abstract}
We study two-point event shapes in planar $\mathcal{N}=4$ super Yang--Mills, with source, sink and detectors given in terms of half-BPS scalars and their light transforms. Akin to local correlation functions, at order by order in perturbation theory, in the large charge limit, this event shape factorizes into the product of two octagons evaluated on the unit circle. Borrowing the octagon results, we study its behavior with respect to the coupling and the angle between the detectors.
\end{abstract}

\maketitle

\section{Introduction}
Event shapes are the natural Lorentzian observables of a conformal field theory~\cite{Hofman:2008ar,Moult:2025nhu}.  They are built from light-ray operators~\cite{Kravchuk:2018htv,Kologlu:2019bco,Belitsky:2013xxa,Belitsky:2013bja} and describe correlations between detectors surrounding a source. They are also hard. For example, the energy--energy correlation in both QCD~\cite{Dixon:2018qgp,DelDuca:2016csb} and $\mathcal{N}=4$ super Yang--Mills theory is known only to the first few perturbative orders~\cite{Engelund:2012re,Belitsky:2013ofa,Henn:2019gkr}. While the conformal bootstrap has enabled numerical progress~\cite{Dempsey:2025yiv}, analytic control at finite coupling remains limited.

For local correlators in $\mathcal{N}=4$ sYM, one way to gain such analytic control is to make the operators heavy. The same is true for event shapes. For example, the energy--energy correlator in the presence of heavy sources~\cite{Chicherin:2023gxt,Firat:2023lbp,Cuomo:2025pjp} was recently computed at finite coupling~\cite{Grassi:2026tdb,Georgoudis:2026vjm}.

Here we consider a different event shape in which the source, sink and detectors are equally heavy. For local correlators, this is the large-charge limit of the \emph{simplest correlator}~\cite{Coronado:2018ypq,Coronado:2018cxj}, which factorizes at every order in perturbation theory into the square of a more fundamental object: the \emph{octagon}, known at all couplings~\cite{Kostov:2019stn,Kostov:2019auq,Belitsky:2019fan,Belitsky:2020qrm}.

We refer to our large-charge event shape as the \emph{simplest event shape}, not only by analogy with~\cite{Coronado:2018ypq} but also because the light transform simplifies. The transform, which usually averages the local correlator over the Lorentzian region, collapses at large charge and localizes the correlator at a single point. Thus, like the local correlator, the large-charge event shape factorizes into the square of a simpler object we call the \emph{octagon event shape}.

Our main result is~\eqref{eq:mainResult}. We show that, order by order in perturbation theory, the octagon event shape is proportional to the octagon correlator evaluated on the unit circle at a phase that encodes the angle between the detectors. We use the exact results~\cite{Kostov:2019stn,Kostov:2019auq,Belitsky:2020qrm} to propose a finite-coupling formulation of the octagon event shape. We study this proposal at weak, finite and strong coupling and in the collinear and back-to-back limits.

\section{The observable}

We work in $\mathcal{N}=4$ sYM in the planar limit, with fixed 't Hooft coupling $\lambda=g_{\mathrm{YM}}^{2}N/(16\pi^{2})$.  The
operators we consider are the ubiquitous single-trace half-BPS scalars
\begin{equation}
  O_k(x) = \operatorname{Tr}\left[(y \cdot \Phi(x))^{2k}\right], \quad y \cdot y = 0
\end{equation}
and the detectors are their light transform~\cite{Kravchuk:2018htv},
\begin{equation}
  \mathcal{O}_k(n)
  =\int_{-\infty}^{\infty}d\tau\,
  \lim_{r\to\infty}r^{2k}\,
  O_k\!\left(t=\tau+r,\;r\vec{n}\right),
  \label{eq:light-transform}
\end{equation}
where $n=(1,\vec{n})$ is the null vector along the detector direction. Replacing the scalar insertion by the appropriate projection of the stress tensor gives the calorimeter considered in~\cite{Hofman:2008ar,Henn:2019gkr}. Unlike the physical calorimeter, generic scalar detectors do not measure a conserved
flux and are therefore not subject to the positivity constraints of the averaged null energy condition~\cite{Faulkner:2016mzt,Hartman:2016lgu}, which constrain energy correlators~\cite{Mecaj:2025ecl,Mecaj:2026kji}. Nonetheless, they are well-defined non-local operators of the theory, whose correlation functions remain largely unexplored.

\begin{figure}[t]
  \centering
  \begin{tikzpicture}[
    every node/.style={font=\small},
    ray/.style={blue!65!black,line width=0.9pt,-{Stealth[length=2mm,width=1.3mm]}},
    bridge/.style={blue!65!black,line width=0.9pt},
    empty bridge/.style={gray!65,line width=0.45pt,dash pattern=on 2pt off 2pt},
    source vertex/.style={circle,draw=black,fill=black,minimum size=4.8pt,inner sep=0pt},
    detector vertex/.style={circle,draw=blue!65!black,fill=white,line width=0.8pt,minimum size=5.8pt,inner sep=0pt}
  ]
    \draw[gray!65,line width=0.45pt,dash pattern=on 2pt off 2pt] (0,0) circle (1.2);
    \draw[ray] (0,0)--(50:1.10);
    \draw[ray] (0,0)--(130:1.10);
    \draw[line width=0.5pt] (50:0.46) arc (50:130:0.46);
    \node at (0,0.25) {$\theta$};
    \node at (28:0.85) {$n$};
    \node at (152:0.85) {$n'$};
    \node[source vertex] at (0,0) {};
    \node[below=3pt,font=\footnotesize] at (0,0) {source};
    \node[detector vertex] at (50:1.2) {};
    \node[detector vertex] at (130:1.2) {};
    \node[above right=3pt] at (50:1.2) {$\mathcal{O}_k$};
    \node[above left=3pt] at (130:1.2) {$\mathcal{O}_k$};
    \node at (0,-1.85) {(a)};
    \begin{scope}[xshift=4.2cm]
      \coordinate (p1) at (-1.05,1.05);
      \coordinate (p2) at (1.05,1.05);
      \coordinate (p3) at (-1.05,-1.05);
      \coordinate (p4) at (1.05,-1.05);
      \draw[empty bridge] (p1)--(p4);
      \draw[empty bridge] (p2)--(p3);
      \draw[bridge] (p1)--(p2) node[midway,above=2pt,text=black] {$k$};
      \draw[bridge] (p3)--(p4) node[midway,below=2pt,text=black] {$k$};
      \draw[bridge] (p1)--(p3) node[midway,left=2pt,text=black] {$k$};
      \draw[bridge] (p2)--(p4) node[midway,right=2pt,text=black] {$k$};
      \node[source vertex] at (p1) {};
      \node[source vertex] at (p4) {};
      \node[detector vertex] at (p2) {};
      \node[detector vertex] at (p3) {};
      \node[above left=3pt] at (p1) {$O_1$};
      \node[above right=3pt] at (p2) {$O_2$};
      \node[below left=3pt] at (p3) {$O_3$};
      \node[below right=3pt] at (p4) {$O_4$};
      \node at (0,-1.85) {(b)};
    \end{scope}
  \end{tikzpicture}
  \caption{(a) The two scalar detectors lie on null infinity in the directions $n$ and $n'$, separated by the angle $\theta$.  (b) The R-symmetry component has $k$ propagators on each perimeter edge and none on the diagonals. Filled vertices denote the source and sink; open vertices denote the detectors. The explicit field realisation is constructed in~\cite{Coronado:2018cxj,Coronado:2018ypq}.}
  \label{fig:setup}
\end{figure}
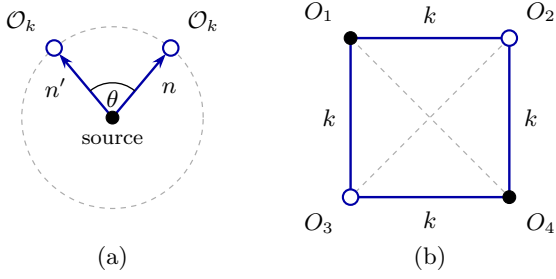

The observable we consider is the Wightman correlation function in Fig.~\ref{fig:setup}(a), of two scalar detectors in the presence of a source and sink of the same dimension
\begin{equation}
  \langle \mathcal{O}_k(n)\mathcal{O}_k(n')\rangle_{x}
  = \langle 0 | O_k(x)\mathcal{O}_k(n)
  \mathcal{O}_k(n')O_k(0)|0\rangle,
  \label{eq:detector-correlator}
\end{equation}
and it is fixed by conformal symmetry to have the form
\begin{equation}
  \langle\mathcal{O}_k
  \mathcal{O}_k\rangle_{x}
  =-\frac{4\pi^{5}\gamma}{k}\bigl[((xn)-i\epsilon)
  ((xn')-i\epsilon)\bigr]^{1-2k}\mathcal{G}_k(\gamma)\label{eq:event-shape}
\end{equation}
where $\mathcal{G}_k(\gamma)$ is the event shape at generic dimension and $\gamma$ is the conformally invariant combination
\begin{equation}
  \gamma=\frac{2\left((xn)-i\epsilon\right)\left((xn')-i\epsilon\right)}
  {\left(x^{2}-i\epsilon x^{0}\right)(nn')}.
  \label{eq:gamma-variable}
\end{equation}

In the rest frame of the source--sink separation $x$, one can define $\gamma=1/\sin^{2}(\theta/2)$, where $\theta$ is the angle between the two detectors, running over $0\leq\theta\leq\pi$ from the collinear to the back-to-back configuration.

While we consider event shapes of arbitrary dimensions, or equivalently arbitrary $R$-charge $k$, our main focus will be the large-charge limit. As we show, order by order in perturbation theory it factorizes as
\begin{equation}
  \lim_{k\to\infty}\mathcal{G}_k(\gamma) = \mathbb{G}(\gamma)^2,\label{eq:def-heavy}
\end{equation}
where $\mathbb{G}(\gamma)$ is what we call the octagon event shape.

\vspace{0.5em}
\noindent\textbf{From correlators to event shapes.}
We start by defining the connected Euclidean four-point function
\begin{equation}
  \langle
  O_1(x_1)O_2(x_2)O_3(x_3)O_4(x_4)
  \rangle_{\mathrm c}
  =\frac{G_k(u,v)}
  {x_{12}^{2k} x_{13}^{2k} x_{24}^{2k} x_{34}^{2k}}
  \label{eq:reduced-correlator}
\end{equation}
where the conformal cross-ratios are given by
\begin{equation}
  u = z\bar z=\frac{x_{12}^{2}x_{34}^{2}}{x_{14}^{2}x_{23}^{2}},
  \qquad
  v = (1-z)(1-\bar z)=\frac{x_{13}^{2}x_{24}^{2}}{x_{14}^{2}x_{23}^{2}}.
  \label{eq:cross-ratios}
\end{equation}

As usual, the analytic continuation from the Euclidean to the Wightman correlator is conveniently performed in Mellin space~\cite{Mack:2009mi,Penedones:2010ue,Rastelli:2016nze,Rastelli:2017udc,Belitsky:2013xxa}. Appendix A of~\cite{Belitsky:2013xxa} already establishes this construction for event shapes of arbitrary dimensions. The only difference is that here we work directly in position-space, where~\eqref{eq:event-shape} reads
\begin{align}
  &\mathcal{G}_k(\gamma) = \int \frac{dj_1}{2\pi i} \int \frac{dj_2}{2\pi i} K_k(j_1,j_2) M_k(j_1,j_2) \gamma^{j_{1}+j_{2}-1}, \nonumber \\
  &K_k(j_1,j_2) = \frac{k}{\pi^{3}2^{4k}} \left(\frac{\Gamma(2k-1-j_{1}-j_{2})}
  {\Gamma(k-j_{1})\Gamma(k-j_{2})}\right)^{2},\label{eq:mellin-master-formula}
\end{align}
where $M_k$ is the Mellin transform of the correlator $G_k$ and the contours run parallel to the imaginary axes with $\operatorname{Re}(j_1+j_2)<2k-1$.

To express the event shape directly in terms of the local correlator~\eqref{eq:reduced-correlator}, we follow the approach of~\cite{Belitsky:2013xxa,Henn:2019gkr} and consider discontinuities
\begin{equation}
  \operatorname{Disc}_{x}F(x)
  \equiv\frac{F(x+i0)-F(x-i0)}{2i}.
  \label{eq:discontinuity-convention}
\end{equation}

More precisely, we write the Mellin transform $M_k$ in terms of the mixed discontinuity of the local correlator across $z<0$ and $\bar{z}>1$. Similarly to~\cite{Belitsky:2013xxa}, the discontinuities generate extra terms that balance the exponential behavior of the kernel $K_k$ at large imaginary values of $j_i$ and allow us to integrate over the Mellin parameters, see Appendix~\ref{ap:detector-transform}. The outcome is that the event shape can be written as
\begin{align}
  \mathcal{G}_k&(\gamma) = \int_{-\infty}^{0}\!\!dz\int_{1}^{\infty}\!\!d\bar z\;
  \mathcal{K}_{k}\operatorname{Disc}_{z}\operatorname{Disc}_{\bar z}G_{k}(z,\bar z),\label{eq:mixed-discontinuity-kernel} \\
  \mathcal{K}_k& = \frac{k(z-\bar z)\gamma^{2k-2}}{\pi^5\,16^k(uv)^k} \,\mathcal{R}_{u,k}\mathcal{R}_{v,k} \left[\frac{1}{\sqrt{(\gamma-u-v)^2-4uv}}\right],\nonumber
\end{align}
where $u = u(z,\bar{z})$ and $v=v(z,\bar{z})$ according to~\eqref{eq:cross-ratios}. 

The $\mathcal{R}_{w,k}$ are differential operators, known as Taylor remainder operators, which build the kernel for generic values of charge $k$ and are defined as 
\begin{align}
  \mathcal{R}_{w,k}f(u,v) &= f(u,v)-\sum_{p=0}^{k-1}\frac{w^p}{p!}\,\left(\partial_w^p f(u,v)\right)_{w=0}.
\end{align}

Below we use both~\eqref{eq:mellin-master-formula} and~\eqref{eq:mixed-discontinuity-kernel} to evaluate the event shape $\mathcal{G}_k$ at generic values of charge $k$ in perturbation theory. We then take $k \to \infty$ and generate perturbative data for the large charge event shapes~\eqref{eq:def-heavy}, which will serve as a check of our all-loop proposal~\eqref{eq:mainResult}.

\section{Weak coupling}\label{sec:weak}
We select the R-symmetry component as depicted in Fig.~\ref{fig:setup}(b), copying the configuration of the simplest correlator~\cite{Coronado:2018ypq,Coronado:2018cxj}. This choice is designed to eliminate the direct cross-talk between the two scalar detectors that gives divergent contributions to the event shape~\cite{Belitsky:2013xxa,Belitsky:2013bja,Caron-Huot:2022eqs}.

We expand the event shape in perturbation theory as
\begin{equation}
  \mathcal{G}_k=\sum_{\ell=0}^{\infty}\lambda^\ell\mathcal{G}_k^{(\ell)},
  \label{eq:weak-coupling-expansion-finite-k}
\end{equation}
and compute the $\ell=0,1,2$ terms below.

While~\eqref{eq:mellin-master-formula} and~\eqref{eq:mixed-discontinuity-kernel} are used to evaluate $\ell >0$ perturbative event shapes~\eqref{eq:weak-coupling-expansion-finite-k}, neither formulation applies at $\ell=0$. At leading order the local correlator is a constant, which admits no Mellin representation and clearly has vanishing mixed discontinuity. Nonetheless, the Wightman correlator~\eqref{eq:detector-correlator} still holds, and the light transforms~\eqref{eq:light-transform} can be trivially evaluated at this order
\begin{equation}
  \int\frac{d\tau}{\left[(\tau-i\epsilon)\left((xn)-\tau-i\epsilon\right)\right]^{k}}
  =\frac{2\pi i \,\Gamma(2k-1)}{\Gamma(k)^2\left((xn)-i\epsilon\right)^{2k-1}},
\end{equation}
so that the leading order event shape reads
\begin{equation}
  \mathcal{G}_k^{(0)}(\gamma)
  =\frac{k}{\pi^3 2^{4k}\gamma}
  \left[\frac{\Gamma(2k-1)}{\Gamma(k)^2}\right]^2.
  \label{eq:finite-k-tree-level-event-shape}
\end{equation}

At this order the detectors are uncorrelated, and all the angular dependence of $\mathcal{G}_k^{(0)}$ is kinematic. The genuine correlation between detectors starts at one loop, which we address below. Nonetheless, at leading order one can still consider the large-charge limit which reads
\begin{equation}
  \lim_{k\to\infty}\mathcal{G}_k^{(0)}
  =\frac{1}{16\pi^4\gamma}
  =\frac{\sin^2(\theta/2)}{16\pi^4}.
  \label{eq:heavy-tree-level-event-shape}
\end{equation}

The first perturbative correction is the same for all charges $k>1$, and it is given by
\begin{equation}
  M_k^{(1)}(j_1,j_2)
  =-2\left(
    \Gamma(-j_1)\Gamma(-j_2)\Gamma(1+j_1+j_2)
  \right)^2
  .
  \label{eq:one-loop-box-mellin-amplitude}
\end{equation}
By plugging this Mellin representation in~\eqref{eq:mellin-master-formula} one obtains the one-loop event shape for arbitrary $k>1$
\begin{equation}
  \mathcal{G}_k^{(1)}(\gamma)
   =-\frac{k}{2^{4k-2}\pi^3\gamma}
  \sum_{r=0}^{2k-2}c_{k,r}\gamma^r B_r(\gamma)\,,
  \label{eq:finite-k-one-loop-event-shape}
\end{equation}
where
\begin{align}
  c_{k,r}&=\sum_{m=0}^{k-1}
  \binom{r}{m}^{2}
  \binom{2k-2-r}{k-1-m}^{2}
  \label{eq:one-loop-ckr}\\
  B_r(\gamma) &= \int_{0}^{1/\gamma} \frac{t^r \log t}{t-1} dt = \zeta_2 -\operatorname{Li}_2\left(1-\frac{1}{\gamma}\right)\label{eq:one-loop-Br}\\
  &-\sum_{m=1}^{r}\gamma^{-m}\left(\frac{1}{m^2}+\frac{\log \gamma}{m}\right).\nonumber
\end{align}

The Mellin representation for $k=1$ is supplemented by additional terms. This is because our choice of R-charge polarization is a mixture of two $SU(4)$ representations: $G_1^{(1)}=(2/3)G_{\mathbf{105}}^{(1)}-(2/3)G_{\mathbf{84}}^{(1)}$; see Appendix~\ref{ap:k1-event-shape}. It turns out that at one loop the two channels integrate to proportional event shapes, $\mathcal{G}_{\mathbf{84}}^{(1)}=-(1/2)\mathcal{G}_{\mathbf{105}}^{(1)}$, and hence

\begin{equation}
  \mathcal{G}_1^{(1)}(\gamma) = \mathcal{G}_{\mathbf{105}}^{(1)}
  =\frac{\operatorname{Li}_2(1-1/\gamma)-\zeta_2}
  {4\pi^3\gamma}.
  \label{eq:k-one-one-loop-event-shape}
\end{equation}
This is the same as the pure $\mathbf{105}$ projection considered in~\cite{Belitsky:2013xxa}. Beyond one loop, these two representations no longer integrate to proportional event shapes; see Appendix~\ref{ap:k1-event-shape}. In fact, the event shape associated with representation $\mathbf{84}$ diverges, while the one of $\mathbf{105}$ remains finite---as it should, since it is precisely this component that gets related to the conserved energy correlator via the supersymmetric Ward identities~\cite{Belitsky:2013xxa,Belitsky:2014zha}.

Even though conservation arguments do not apply to generic scalar event shapes~\eqref{eq:event-shape}, in the next section we show that it is indeed possible to define finite scalar event shapes at any given order in perturbation theory. One example is the large charge event shape~\eqref{eq:def-heavy}, which is obtained by resumming~\eqref{eq:finite-k-one-loop-event-shape} and reads

\begin{equation}
  \lim_{k\to \infty}\mathcal{G}_k^{(1)}
  =-\frac{\sin^3(\theta/2)\operatorname{Cl}_2(\theta)}
  {2\pi^4\cos(\theta/2)}\,, \label{eq:large-k-one-loop-event-shape}
\end{equation}
where $\operatorname{Cl}_2(\theta)$ is one of the Clausen functions
\begin{equation}
  \operatorname{Cl}_{2\ell}(\theta)
  =\operatorname{Im}\operatorname{Li}_{2\ell}(e^{i\theta}). \label{eq:Clausen-functions}
\end{equation}

One can also perform an analogous computation at two loops, see Appendix~\ref{ap:two-loop-detectors}, and find that the next perturbative correction of the large charge event shape is
\begin{equation}
  \lim_{k\to \infty} \mathcal{G}_k^{(2)}
  =\frac{3\sin^{4}(\theta/2)}{\pi^{4}\cos^{2}(\theta/2)}
  \left[\frac{\operatorname{Cl}_{2}(\theta)^{2}}{3}+\frac{\operatorname{Cl}_{4}(\theta)}{\tan(\theta/2)}\right].
  \label{eq:large-k-two-loop-event-shape}
\end{equation}

Here at tree level~\eqref{eq:heavy-tree-level-event-shape}, one loop~\eqref{eq:large-k-one-loop-event-shape} and two loops~\eqref{eq:large-k-two-loop-event-shape}, we have first performed the detector integrals and then taken the charge to be large. Below we argue that it is possible to do it in the opposite order.

\section{The simplest event shape}
At every order in perturbation theory, the large-charge local correlator factorizes as~\cite{Coronado:2018ypq,Coronado:2018cxj},
\begin{equation}
  \lim_{k\to\infty}G_k(z,\bar z)=\mathbb{O}(z_{\text{o}},\bar{z}_{\text{o}})^2,
  \label{eq:octagon-limit}
\end{equation}
where $\mathbb{O}$ is the \emph{octagon}, defined in terms of the cross-ratios $z_{\text{o}} = z/(z-1)$ and $\bar{z}_{\text{o}} = \bar z/(\bar z-1)$.

Order by order in the coupling, the octagon is given by a combination of conformal ladder integrals~\cite{Usyukina:1993ch,Coronado:2018ypq,Coronado:2018cxj},
\begin{equation}
  \mathbb O^{(\ell)}(z_{\text{o}},\bar{z}_{\text{o}})
  =\sum_{m\geq1}\ \sum_{\ell_1+\dots+\ell_m=\ell}
  c_{\ell_1\cdots\ell_m}\prod_{i=1}^{m}\frac{f_{\ell_i}(z,\bar z)}{z-\bar z},
  \label{eq:octagon-ladders}
\end{equation}
with rational coefficients $c_{\ell_1\cdots\ell_m}$ and pure polylogarithms $f_{L}$ of weight $2L$ in the alphabet $\{z,1-z,\bar z,1-\bar z\}$. Apart from the powers of $1/(z-\bar z)$, no other rational function of the cross-ratios appears.

Let's assume that the large-charge limit commutes with the detector integrals, so that we can take $k\to\infty$ directly at the level of the integrand in~\eqref{eq:mixed-discontinuity-kernel}. As detailed in Appendix~\ref{ap:limits-interchange}, the detector kernel simplifies to
\begin{equation}
  \lim_{k\to\infty}\mathcal{K}_{k}(z,\bar z;\gamma)
  =\frac{z-\bar z}{\pi^{6}\gamma(\gamma-4u)(\gamma-4v)},
  \label{eq:rational-kernel}
\end{equation}
so the event shape can be written as
\begin{equation}
  \lim_{k\to\infty}\mathcal{G}_k
  =
  -\int_{-\infty}^{0}\hspace{-0.3em}du\int_{-\infty}^{0}\hspace{-0.3em}dv\,
  \frac{\operatorname{Disc}_{z}\operatorname{Disc}_{\bar z}\mathbb{O}(z_{\text{o}},\bar{z}_{\text{o}})^2}{\pi^6\gamma(\gamma-4u)(\gamma-4v)}.
  \label{eq:heavy-dispersion}
\end{equation}

The light transform has become a double dispersion relation. Each integral is a Cauchy integral along a single cut which reconstructs the squared octagon at the point $u=v=\gamma/4$. In terms of the cross-ratios $\{z_{\text{o}},\bar{z}_{\text{o}}\}$, this point lies beautifully on the unit circle, so the large-charge event shape, i.e. the simplest event shape, becomes
\begin{equation}
  \lim_{k\to\infty}\mathcal{G}_k
  =\frac{\sin^{2}(\theta/2)}{16\pi^{4}}
  \mathbb{O}(e^{-i\theta},e^{i\theta})^{2},
  \label{eq:mainResult1}
\end{equation}
and the octagon event shape~\eqref{eq:def-heavy} is
\begin{equation}
  \mathbb{G}(\theta)
  =\frac{\sin(\theta/2)}{4\pi^{2}}
  \mathbb{O}(e^{-i\theta},e^{i\theta}).
  \label{eq:mainResult}
\end{equation}

To conclude this derivation, we address two aspects of scalar event shapes we have postponed so far: their finiteness in perturbation theory and the exchange of the detector integral with the large charge limit.

\vspace{0.5em}
\noindent\textbf{Finiteness.} Eliminating detector cross-talk is necessary but not sufficient for convergence~\cite{Belitsky:2013xxa}. What we argue below is that at any fixed order $\ell$ in perturbation theory, there is a finite charge $k_\ast(\ell)$ such that the event shape~\eqref{eq:event-shape} is finite for all $k\geq k_\ast(\ell)$. The central assumption is that correlation functions saturate in perturbation theory:\footnote{Using the ten-dimensional integrand of~\cite{Caron-Huot:2021usw}, we find $k_\ast(\ell)=\ell+1$ through seven loops. Hexagonalization gives the more conservative bound $k_\ast(\ell)\leq\ell+m_\ell$, where $m_\ell$ is the largest number of mirror particles that can appear at $\ell$ loops~\cite{Coronado:2018ypq}, e.g. $m_5=3$.}
\begin{equation}
  G_k^{(\ell)}(z,\bar{z}) = \left[\mathbb O\left(z_{\text{o}},\bar{z}_{\text{o}}\right)^2\right]^{(\ell)},\quad  \text{for} \quad k\geq k_\ast(\ell).
  \label{eq:finite-charge-octagon-window}
\end{equation}

The reasoning is the same as for the octagon: a planar diagram that crosses bridges of width $k$ costs $k$ powers of the coupling, so at fixed loop order and large enough charge the diagrams that would couple the ``inside'' and the ``outside'' are suppressed, and the correlator factorizes into the square of a more fundamental object, the octagon~\cite{Coronado:2018ypq,Caron-Huot:2021usw}. Thus, for $k\geq k_\ast(\ell)$, local correlators are also given by the combination of ladders~\eqref{eq:octagon-ladders}.

As recalled in Appendix~\ref{ap:k1-event-shape}, the two-loop divergence at $k=1$ arises from the Regge limit $|u|,|v|\to\infty$ with ratio fixed. In the detector integral~\eqref{eq:mixed-discontinuity-kernel}, this corresponds to $z=-R\cos\phi$ and $\bar z=R\sin\phi$, with $R\to\infty$ at fixed $0<\phi<\pi/2$. However, if a correlator is given by the combination of ladders~\eqref{eq:octagon-ladders}, then it does not diverge: the factor $1/|z-\bar z|$ always decays, the mixed discontinuity of ladders $f_{L}$ is again a sum of products of polylogarithms of real arguments, and the detector kernel \eqref{eq:mixed-discontinuity-kernel} and radial measure supply another charge-independent factor of $O(R^{-2})$. Thus at most the event shape scales as $\int dR R^{-2}\log^{p}R$ and converges absolutely.

This statement holds order by order in the coupling, and it does not address finiteness below the threshold $k_{\ast}(\ell)$. For example, the case $k=1$ is below the threshold at one loop and converges, while at two loops it is again below threshold but diverges. 

\vspace{0.5em}
\noindent\textbf{Order of limits.} The argument again relies on saturation. At any given order $\ell$ in perturbation theory, once $k\geq k_\ast(\ell)$, the correlator saturates to the octagon~\eqref{eq:finite-charge-octagon-window}. So the large $k$ limit of the integrand in~\eqref{eq:mixed-discontinuity-kernel} is governed only by the kernel $\mathcal{K}_k$.

As shown in Appendix~\ref{ap:detector-transform}, the full detector integrand admits a $k$-independent integrable bound~\eqref{eq:appendix-full-majorant}. Dominated convergence therefore justifies exchanging the large-charge limit with the detector integrals at any given order in perturbation theory and any fixed interior angle.

\section{Properties}

\vspace{0.5em}
\noindent\textbf{Weak coupling.} Expanding the octagon at weak coupling in~\eqref{eq:mainResult} gives the perturbative octagon event shape
\begin{equation}
  \mathbb{G}(\theta)
  \overset{\lambda \to 0}{\sim}\frac{\sin(\theta/2)}{4\pi^{2}}
  \left(1-\lambda q_{1}+\frac{\lambda^{2} q_2}{2}
  -\frac{\lambda^{3}q_3}{6}+O(\lambda^{4})\right),
  \label{eq:weak-expansion}
\end{equation}
where the $q_\ell$ are given in terms of Clausen functions
\begin{equation}
  q_{\ell}(\theta) = 2(2\ell)!\tan(\theta/2)\operatorname{Cl}_{2\ell}(\theta).
  \label{eq:weak-moments}
\end{equation}

Squaring and expanding~\eqref{eq:weak-expansion} according to~\eqref{eq:def-heavy} reproduces the tree-level~\eqref{eq:heavy-tree-level-event-shape}, one-loop~\eqref{eq:large-k-one-loop-event-shape} and two-loop~\eqref{eq:large-k-two-loop-event-shape} results of the large charge event shape. Each expansion was obtained by taking detector and large charge limits in a different order, and their agreement exemplifies the commutation of these limits.

\vspace{0.5em}
\noindent\textbf{Finite coupling.} The octagon can be written as a Fredholm determinant of a modified Bessel kernel at finite coupling~\cite{Kostov:2019stn,Kostov:2019auq,Belitsky:2020qrm}. We propose that the finite-coupling octagon event shape~\eqref{eq:mainResult} follows the same completion
\begin{equation}
  \mathbb{G}(\theta)
  =\frac{\sin(\theta/2)}{4\pi^{2}}
  \det\left(1-\mathbb{K}\right),
  \label{eq:determinant-master}
\end{equation}
where for $n,m\geq 0$, the matrix elements are given by
\begin{align}
  \mathbb{K}_{nm}
  ={}&(-1)^{n+m}2(2n+1)\times
  \nonumber\\
  &\times\int_{0}^{\infty}\frac{dt}{t}\,
  J_{2m+1}(t)\, \chi\left(\tfrac{t}{2\sqrt{\lambda}};\theta\right)J_{2n+1}(t).
  \label{eq:bessel-kernel}
\end{align}
Both the angle and the coupling dependence enter solely in $\chi(t,\theta)$, the symbol of the Bessel kernel~\cite{Belitsky:2020qrm,Belitsky:2020qir,Bajnok:2024ymr}
\begin{equation}
  \chi(t;\theta)=\frac{\sin^{2}(\theta/2)}{\sin^{2}(\theta/2)+\sinh^{2}(t/2)}.
  \label{eq:symbol}
\end{equation}

\begin{figure}[t]
  \centering
  \begin{tikzpicture}
    \begin{axis}[
      width=0.97\columnwidth, height=0.66\columnwidth,
      xmin=0, xmax=pi, ymin=-0.04/(4*pi^2), ymax=1.04/(4*pi^2),
      xtick={0,pi/4,pi/2,3*pi/4,pi},
      xticklabels={$0$,$\frac{\pi}{4}$,$\frac{\pi}{2}$,$\frac{3\pi}{4}$,$\pi$},
      ytick={0,0.01,0.02},
      yticklabels={$0$,$0.01$,$0.02$},
      scaled y ticks=false,
      xlabel={$\theta$},
      ylabel={$\mathbb{G}(\theta)$},
      tick label style={font=\footnotesize},
      label style={font=\small},
      legend style={font=\scriptsize, draw=none, fill=none, at={(0.02,0.98)}, anchor=north west},
      legend image post style={xscale=1.1},
      legend cell align=left,
      every axis plot/.append style={thick, no markers},
    ]
      \addplot[blue!85!black]    table[x=theta, y expr={sin(deg(\thisrow{theta})/2)/(4*pi^2)*sqrt(1-\thisrow{lam0p5})}] {finite-coupling.dat};
      \addlegendentry{$\lambda=1/16$}
      \addplot[orange]          table[x=theta, y expr={sin(deg(\thisrow{theta})/2)/(4*pi^2)*sqrt(1-\thisrow{lam2})}] {finite-coupling.dat};
      \addlegendentry{$\lambda=1/4$}
      \addplot[green!60!black]   table[x=theta, y expr={sin(deg(\thisrow{theta})/2)/(4*pi^2)*sqrt(1-\thisrow{lam8})}] {finite-coupling.dat};
      \addlegendentry{$\lambda=1$}
      \addplot[red!85!black]     table[x=theta, y expr={sin(deg(\thisrow{theta})/2)/(4*pi^2)*sqrt(1-\thisrow{lam32})}] {finite-coupling.dat};
      \addlegendentry{$\lambda=4$}
      \addplot[black, dashed, domain=0:pi, samples=100] {0};
      \addplot[black, dotted, domain=0:pi, samples=100] {sin(deg(x)/2)/(4*pi^2)};
    \end{axis}
  \end{tikzpicture}
  \caption{The octagon event shape at finite coupling. It is bounded from above by tree level (dotted line) and from below by strong coupling (dashed line). The Bessel matrix is truncated at size $64\times64$.}
  \label{fig:finite-coupling}
\end{figure}
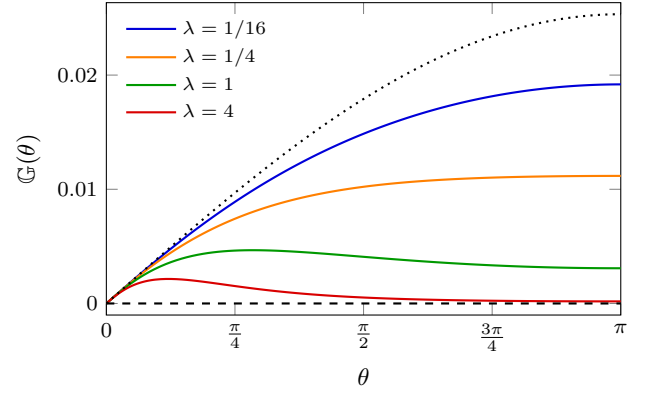

At weak coupling, the matrix elements of the Bessel kernel~\eqref{eq:bessel-kernel} can be easily expanded and one recognizes that the $q_\ell$ appearing in the perturbative series~\eqref{eq:weak-expansion} are the moments of the symbol~\eqref{eq:symbol}
\begin{equation}
  q_\ell(\theta)=2\ell\int_0^\infty\!dt\,t^{2\ell-1}\chi(t;\theta).
  \label{eq:symbol-moments}
\end{equation}

At finite coupling, the one-dimensional integrals in~\eqref{eq:bessel-kernel} can be evaluated numerically for any value of the angle between the detectors. By considering larger and larger determinants in~\eqref{eq:determinant-master} one can obtain the octagon event shape for any value of the coupling, as shown in Fig.~\ref{fig:finite-coupling}.

\vspace{0.5em}
\noindent\textbf{Strong coupling.} The asymptotics of the Bessel determinant are known at strong coupling~\cite{Bargheer:2019exp,Belitsky:2020qrm,Belitsky:2020qir,Bajnok:2024ymr} and consequently so is the octagon~\cite{Bargheer:2019exp}. Thus, the strong-coupling expansion of the octagon event shape reads
\begin{align}
  \mathbb{G}(\theta)&\overset{\lambda \to \infty}{\sim}\frac{\sin(\theta/2)}{4\pi^{2}}\lambda^{1/4}\,
  e^{\frac{\theta(\theta-2\pi)}{\pi}\sqrt{\lambda}+B(\theta)}\left(1-\frac{3I_1(\theta)}{16\sqrt{\lambda}}\right.\nonumber\\
  &
  \left. -\frac{15I_1(\theta)^2}{512\lambda}+\frac{40I_2(\theta)-65I_1(\theta)^3}{8192\lambda^{3/2}}+O(\lambda^{-2})\right)\label{eq:strong-coupling-event-shape}  
\end{align}
where the corrections are also controlled by the symbol~\eqref{eq:symbol} through the integrals~\cite{Belitsky:2020qrm,Belitsky:2020qir,Bajnok:2024ymr}
\begin{align}
  I_\ell(\theta)&=\int_0^\infty\!dt\,
  \frac{(t^{-1}\partial_t)^\ell}{\pi(2\ell-1)!!}\left(t\partial_t\log\bigl(1-\chi(t;\theta)\bigr)\right),
  \label{eq:strong-coupling-profile}\\
  B(\theta)&=\frac12\log(2\pi)
  +\frac12\int_0^\infty ds\,
  \left(s\psi_\theta(s)^2-\frac{1-e^{-s}}{s}\right),\nonumber
  \\
  \psi_\theta(s)&=\int_0^\infty\frac{dt}{\pi}\,
  \cos(st)\log\bigl(1-\chi(t;\theta)\bigr).\label{eq:strong-coupling-constant}
\end{align}

For every fixed physical angle away from the endpoints ($0 < \theta < \pi$), the octagon event shape at strong coupling is exponentially suppressed just like the octagon~\cite{Bargheer:2019exp}, and it approaches zero from above.

\vspace{0.5em}
\noindent\textbf{Probability.} In the integrability description, the octagon is built by joining two hexagons~\cite{Basso:2015zoa,Fleury:2016ykk} and summing over virtual mirror particles and their bound states. In the determinant~\eqref{eq:determinant-master}, the bound states are resummed into the symbol, and each mirror particle contributes one factor of this symbol~\cite{Kostov:2019auq}.

For physical detector angles, $0<\chi(t;\theta)\leq1$ can be read as an occupation number. Consider free fermions in the plane with zero angular momentum, where the mode of momentum $t$ is occupied with probability $\chi(t;\theta)$. The octagon event shape (normalized by its tree-level value) is the probability to find no fermion in the disk of radius $2\sqrt{\lambda}$, whose diameter is the Zhukovsky cut~\cite{Soshnikov:2000,Belitsky:2020qir}.

In this picture, $n$ mirror particles correspond to $n$ fermions in the disk. Pauli exclusion makes this cost $\lambda^{n^2}$, which is why $n$ mirror particles first contribute at $n^2$ loops~\cite{Coronado:2018ypq}. At zero coupling the disk has zero size and is certainly empty; increasing the coupling enlarges it and makes emptiness less likely. The octagon event shape therefore decreases monotonically with the coupling and is bounded by its tree-level value~\eqref{eq:weak-expansion} and its strong-coupling limit~\eqref{eq:strong-coupling-event-shape}:
\begin{equation}
  0\leq\mathbb{G}(\theta)\leq\frac{\sin(\theta/2)}{4\pi^{2}}.
  \label{eq:heavy-bound}
\end{equation}

\vspace{0.5em}
\noindent\textbf{Collinear limit.} The octagon event shape vanishes linearly as $\theta\to0$, with a subleading logarithm familiar from energy correlators~\cite{Henn:2019gkr,Dixon:2019uzg},\footnote{Fourier transforming the position-space detector correlator maps the tree-level contribution to a delta function at the back-to-back endpoint~\cite{Belitsky:2013bja}. Thus, in momentum-space the event shape in the collinear limit starts at one loop and scales logarithmically rather than linearly.}
\begin{equation}
  \mathbb{G}(\theta) \overset{\theta \to 0}{\sim} \frac{\theta}{8\pi^2}
  +\lambda\frac{\theta^{3}\log\theta}{4\pi^{2}}
  +O(\theta^3)\,.
  \label{eq:collinear-limit}
\end{equation}
Although higher orders in perturbation theory contribute to regular $O(\theta^3)$ terms in the collinear limit, the coefficient of the leading logarithm is one-loop exact. 

This follows from the ladder expansion~\eqref{eq:octagon-ladders}: higher-loop ladders contribute only non-logarithmic terms at $O(\theta^3)$, while products of ladders contribute to logarithmic terms suppressed by higher powers of $\theta$. The leading logarithm term therefore comes entirely from the single one-loop ladder. This is the detector analogue of the OPE limit of ladder integrals of local correlators~\cite{Belitsky:2020qrm}.

\vspace{0.5em}
\noindent\textbf{Back-to-back limit.} In position space, the angle between detectors is measured in the rest frame of the fixed source--sink separation, while in momentum space the detector angle is defined in the rest frame of their total momentum. Since the Fourier transform integrates over positions, the two endpoints need not probe the same region of the local correlator.\footnote{Event shapes in position and momentum space are closely related in the collinear limit. Since $n\cdot n'\to0$ is frame independent, both probe the detector-pair OPE~\cite{Hofman:2008ar,Kologlu:2019mfz}.}

In our position space formulation, when $\theta\to\pi$, the octagon event shape probes the regular $u=v=1/4$ point of the local correlator and~\eqref{eq:mainResult} reduces to
\begin{equation}
    \mathbb{G}(\theta)\overset{\theta \to \pi}{\sim}\frac{\mathbb{O}(-1,-1)}{4\pi^{2}}.
    \label{eq:position-back-to-back}
\end{equation}

This is vastly different from the momentum-space back-to-back limit of energy correlators, which probes the region $u,v\to0$, where the source and sink become null separated from the detector insertions~\cite{Korchemsky:2019nzm}. These correlators are governed by Sudakov factors and the cusp anomalous dimension~\cite{Korchemsky:2019nzm,Moult:2018jzp}. 

\section{Outlook}
We have shown that, at large charge and order by order in perturbation theory, the light transform evaluates the correlator at a single point. This turns the octagon event shape into the octagon correlator evaluated on the unit circle~\eqref{eq:mainResult}.

The natural next step is the back-to-back limit in momentum space. This regime is more complicated because the angular window around the back-to-back configuration scales with the charge. This requires a subtle double-scaling limit at large charge. We would like to determine whether this regime is controlled by the octagon anomalous dimension $\Gamma_{\mathrm{oct}}$ and related to Coulomb branch amplitudes~\cite{Belitsky:2019fan,Belitsky:2022itf}. This would be akin to the case of energy--energy correlators, whose back-to-back limit is governed by the cusp anomalous dimension $\Gamma_{\mathrm{cusp}}$~\cite{Korchemsky:2019nzm} and related to gluon scattering amplitudes~\cite{Alday:2007hr,Alday:2010zy,Eden:2010zz}.

It would also be interesting to go beyond two detectors. Higher-point correlators of heavy operators are built from similar integrable polygons~\cite{Fleury:2020ykw,Bargheer:2025ybg}, and their light transforms may collapse in the same way. If they do, we could obtain a family of simple and exact multi-detector event shapes whose properties we are just starting to explore.

\subsection{Acknowledgments}
We thank Alessandro Georgoudis, Ian Moult, Vasco Gon\c{c}alves, Pedro Vieira and Alexandre Homrich for insightful discussions. SK thanks Andreas Stergiou for his guidance. The work of C. B. was supported by the European Research Council (ERC) under the European Union's Horizon 2020 research and innovation program (grant agreement No. 865075) EXACTC and by the Science and Technology Facilities Council (STFC) ST/X000753/1. SK is supported by the UK Engineering and Physical Sciences Research Council (EPSRC) through a Doctoral Landscape Award, grant EP/Z535035/1.

\bibliography{references}

\section*{SUPPLEMENTAL MATERIAL}

\appendix
\setcounter{secnumdepth}{2}

\section{The detector transform}
\label{ap:detector-transform}
We begin with the light transform~\eqref{eq:light-transform}. In the detector limit, this reduces to the dimension $\Delta = 2k$ integral evaluated in Appendix A of~\cite{Belitsky:2013xxa},
\begin{equation}
 \begin{aligned}
 &\int_{-\infty}^{\infty}\frac{d\tau}{2\pi}\bigl(\epsilon-i(\tau-(xn))\bigr)^{j_1-k}\bigl(\epsilon+i\tau\bigr)^{j_2-k}=
 \\
 &\quad=\frac{\Gamma(2k-1-j_1-j_2)}{\Gamma(k-j_1)\Gamma(k-j_2)}\bigl(i(xn)+2\epsilon\bigr)^{j_1+j_2+1-2k}.
 \end{aligned}
 \label{eq:appendix-light-transform-integral}
\end{equation}
The Wightman prescription places the two $\tau$ singularities on opposite sides of the real axis, and we retain the regulator through both transforms. The two light transforms then give
\begin{multline}
 -\frac{4\pi^2}{2^{4k}}
 \left[\frac{\Gamma(2k-1-j_1-j_2)}{\Gamma(k-j_1)\Gamma(k-j_2)}\right]^2
 \gamma^{j_1+j_2}
 \\
 \times\bigl[((xn)-i\epsilon)((xn')-i\epsilon)\bigr]^{1-2k},
 \label{eq:appendix-monomial-light-transform}
\end{multline}
which, with our normalization~\eqref{eq:event-shape}, reproduces~\eqref{eq:mellin-master-formula}.

Convergence of~\eqref{eq:appendix-light-transform-integral} at large $|\tau|$ requires $\operatorname{Re}(j_1+j_2)<2k-1$. Absolute convergence suffices to exchange the detector-time and Mellin integrations, and an integrable bound uniform in the Wightman regulator permits its removal inside the Mellin integral.

Continuing $z$ and $\bar z$ independently from the principal Euclidean sheet to $z<0<1<\bar z$, the discontinuity~\eqref{eq:discontinuity-convention} gives
\begin{equation}
  \operatorname{Disc}_{z}\operatorname{Disc}_{\bar z}
  \left[u^{j_1}v^{j_2}\right]=-\sin(\pi j_1)\sin(\pi j_2)
  (-u)^{j_1}(-v)^{j_2}.
  \label{eq:mixed-disc-mode}
\end{equation}

Once we invert the Mellin amplitude in terms of this discontinuity we obtain the kernel in~\eqref{eq:mixed-discontinuity-kernel},
\begin{multline}
 \mathcal{K}_k(z,\bar z;\gamma)
 =
 \int\frac{dj_1}{2\pi i}\int\frac{dj_2}{2\pi i}\,
 \frac{K_k(j_1,j_2)}{\sin(\pi j_1)\sin(\pi j_2)}
 \\
 \times\frac{z-\bar z}{\gamma uv}\left(\frac{\gamma}{-u}\right)^{j_1}
 \left(\frac{\gamma}{-v}\right)^{j_2},
 \label{eq:appendix-kernel-inversion}
\end{multline}
where the contours lie between the adjacent poles with $j_a=-1$ and $j_a=0$ of $1/\sin(\pi j_a)$.

At the end of this appendix we establish the convergence and Mellin--position interchange required to apply~\eqref{eq:appendix-kernel-residues} to saturated correlators, but first we state the result. For saturated correlators, the Mellin contours are chosen within this strip to separate the left and right pole series of the Mellin amplitude. Below saturation, polynomial prefactors can shift the correlator's Mellin contours away from the strip used in~\eqref{eq:appendix-kernel-inversion}. When deforming these contours back into the strip, the residue contributions from any poles crossed must be retained.

Closing both contours to the left for 
\begin{equation}
\sqrt{|u|/\gamma}+\sqrt{|v|/\gamma}<1,
\end{equation} picks the poles $j_1=k-1-p$ and $j_2=k-1-q$, with $p,q\geq k$, giving
\begin{equation}
 \mathcal{K}_k
 =\frac{k(z-\bar z)}{\pi^5 2^{4k}\gamma uv}
 \left(\frac{\gamma^2}{uv}\right)^{k-1}
 \sum_{p,q\geq k}\binom{p+q}{p}^{\!2}
 \frac{u^pv^q}{\gamma^{p+q}}.
 \label{eq:appendix-kernel-residues}
\end{equation}
Applying both Taylor remainder operators to the factor $1/\sqrt{\gamma[(\gamma-u-v)^2-4uv]}$ as in~\eqref{eq:mixed-discontinuity-kernel}, produces the sum~\eqref{eq:appendix-kernel-residues}.

\noindent\textbf{Convergence and interchange.} We establish finiteness by first bounding the detector integral~\eqref{eq:appendix-kernel-inversion} uniformly in $k$. We choose contours $j_a=c+it_a$ with $-1/2<c<0$ fixed independently of $k$ and work under the saturation assumption~\eqref{eq:finite-charge-octagon-window}, so that only the detector kernel depends on $k$. 

On these contours, the Euler integral bounds the numerator by $|\Gamma(2k-1-2c-i(t_1+t_2))|\leq\Gamma(2k-1-2c)$. The reciprocal Gamma functions satisfy
\begin{align}
 \frac{\Gamma(k-c)^2}{|\Gamma(k-c-it)|^2}
 &=\prod_{q=0}^\infty\left(1+\frac{t^2}{(k-c+q)^2}\right)
 \nonumber\\
 &\leq\frac{\Gamma(1-c)^2}{|\Gamma(1-c-it)|^2}.
 \label{eq:appendix-gamma-comparison}
\end{align}
The remaining charge dependence is bounded, since
\begin{align}
 \frac{k\,\Gamma(2k-1-2c)^2}{2^{4k}\Gamma(k-c)^4}
 &=\frac{k}{2^{4+4c}\pi}\left[\frac{\Gamma(k-c-1/2)}{\Gamma(k-c)}\right]^2
 \nonumber\\
 &\longrightarrow\frac{1}{2^{4+4c}\pi}.
 \label{eq:appendix-real-gamma-bound}
\end{align}
Stirling's formula then bounds the kernel's Mellin integrand uniformly in $k$,
\begin{equation}
 \left|\frac{K_k(j_1,j_2)}{\sin(\pi j_1)\sin(\pi j_2)}\right|
 \leq C_c\prod_{a=1}^{2}(1+|t_a|)^{2c-1}.
 \label{eq:appendix-mellin-majorant}
\end{equation}
Since $c<0$, this bound is integrable in both Mellin variables. Taking absolute values in~\eqref{eq:appendix-kernel-inversion} and applying~\eqref{eq:appendix-mellin-majorant} gives
\begin{equation}
 |\mathcal{K}_k(z,\bar z;\gamma)|
 \leq C_{\gamma,c}\frac{|z-\bar z|}{(-u)^{1+c}(-v)^{1+c}},
 \qquad k\geq1,
 \label{eq:appendix-kernel-bound}
\end{equation}
where $C_{\gamma,c}$ is independent of $k,z,\bar z$. The ladder expansion~\eqref{eq:octagon-ladders}, expressed in the detector variables $z,\bar z$, contains only finitely many terms at fixed $\ell$. Every positive-loop term in the squared octagon has at least one factor $1/(z-\bar z)$. Since $\bar z-z>1$ in the integration region, higher inverse powers only improve the bound. In terms of the boundary logarithms
\begin{equation}
 \begin{aligned}
 \mathcal{L}(z,\bar z)&=1+|\log(-z)|+\log(1-z)
 \\
 &\quad+|\log(\bar z-1)|+\log\bar z,
 \end{aligned}
 \label{eq:appendix-boundary-logarithms}
\end{equation}
the polylogarithms $f_L$ of weight $2L$ in~\eqref{eq:octagon-ladders} are bounded on both sides of the cuts $z<0$ and $\bar z>1$ by
\begin{equation}
 \left|f_L(z\pm i0,\bar z\pm i0)\right|
 \leq C_L\,\mathcal{L}(z,\bar z)^{2L}.
 \label{eq:appendix-ladder-growth}
\end{equation}
The mixed discontinuity of each product of ladders is a combination of these boundary values, so
\begin{equation}
 \left|\operatorname{Disc}_z\operatorname{Disc}_{\bar z}
 [\mathbb{O}^2]^{(\ell)}\right|
 \leq C_\ell\frac{\mathcal{L}(z,\bar z)^{2\ell}}{\bar z-z}.
 \label{eq:appendix-ladder-bound}
\end{equation}
The full detector integrand therefore obeys
\begin{multline}
 \left|\mathcal{K}_k\operatorname{Disc}_z\operatorname{Disc}_{\bar z}
 G_k^{(\ell)}\right|
 \\
 \leq C_{\gamma,\ell,c}
 \frac{\mathcal{L}(z,\bar z)^{2\ell}}{[(-z)(1-z)\bar z(\bar z-1)]^{1+c}}.
 \label{eq:appendix-full-majorant}
\end{multline}
In each positive variable $-z$ and $\bar z-1$, this bound has power $-1-c$ near zero and $-2-2c$ at infinity, multiplied by a finite power of logarithms. Since $-1/2<c<0$, it is integrable at every boundary, including simultaneous limits. With the kernel written as the Mellin integral~\eqref{eq:appendix-kernel-inversion}, the same estimates bound the joint integrand by the product of the right-hand sides of~\eqref{eq:appendix-full-majorant} and~\eqref{eq:appendix-mellin-majorant}. Absolute convergence therefore allows us to exchange the kernel's Mellin integrations with the position integrations in~\eqref{eq:mixed-discontinuity-kernel}. In particular, the detector transform is finite at every fixed $k\geq k_\ast(\ell)$.

\section{The \texorpdfstring{$k=1$}{k=1} event shape}
\label{ap:k1-event-shape}

Eliminating detector cross-talk does not by itself ensure finiteness, so the convergence of the detector integrals must be checked separately. We illustrate this with the $k=1$ event shape in perturbation theory. At $k=1$, the correlator is given by a combination of two SU(4) representations only $[2,0,2]=\mathbf{84}$ and $[0,4,0]=\mathbf{105}$ that are simply related
\begin{equation}
  G_{\mathbf{84}}^{(\ell)}=\frac{3u+3v-1}{2}G_{\mathbf{105}}^{(\ell)},
  \qquad \ell\geq1.
  \label{eq:r-k1-channel-relation}
\end{equation}
so that the entire correlator reads
\begin{equation}
  G_1^{(\ell)}=\frac{2}{3}(G_{\mathbf{105}}^{(\ell)}-G_{\mathbf{84}}^{(\ell)})=(1-u-v)G_{\mathbf{105}}^{(\ell)},
  \quad \ell\geq1.
  \label{eq:r-k1-equal-bridge-relation}
\end{equation}
Below we show that at one loop both are finite, while at two loops the $\mathbf{84}$ channel diverges.

\vspace{0.5em}
\noindent\textbf{One loop.} The one-loop $\mathbf{105}$ channel is given by
\begin{equation}
  G_{\mathbf{105}}^{(1)}=-2\Phi^{(1)}(u,v),
  \label{eq:appendix-k1-one-loop-local}
\end{equation}
where the the box integral in the Euclidean region $\bar z=z^*$ can be written as
\begin{equation}
  \Phi^{(1)}(u,v)=-\int_0^1\!dt\,
  \frac{\log(v/u)+2\log t}{vt^2+(1-u-v)t+u}.
  \label{eq:two-loop-box-definition}
\end{equation}

The detector transforms of $u\Phi^{(1)}$ and $v\Phi^{(1)}$ vanish. Hence via~\eqref{eq:r-k1-channel-relation}
\begin{equation}
  \mathcal{G}_{\mathbf{84}}^{(1)} = -\frac{1}{2}\mathcal{G}_{\mathbf{105}}^{(1)}.
  \label{eq:two-loop-k1-one-loop-relation}
\end{equation}

\textbf{Two loops.} The two-loop correlator in the $\mathbf{105}$ representation is~\cite{Belitsky:2013xxa,Belitsky:2013ofa}
\begin{align}
  G_{\mathbf{105}}^{(2)}
  &\!=(1\!+\!u\!+\!v)\bigl(\Phi^{(1)}\bigr)^2
  \!+\!4\bigl(\Phi^{(2)}\!+\!\Phi_u^{(2)}\!+\!\Phi_v^{(2)}\bigr),
  \label{eq:two-loop-105-complete}
\end{align}
where the two-loop ladder is for $\bar z=z^*$ is~\cite{Usyukina:1993ch}
\begin{equation}
  \begin{aligned}
  \Phi^{(2)}(u,v)&=-\frac12\int_0^1\!dt\,
  \frac{\log t}{vt^2+(1-u-v)t+u}
  \\
  &\quad\times[\log\tfrac vu+\log t][\log\tfrac vu+2\log t].
  \end{aligned}
  \label{eq:two-loop-ladder-definition}
\end{equation}
In Euclidean kinematics, the crossed $\ell$-loop ladders are
\begin{equation}
 \Phi_u^{(\ell)}=\frac1u\Phi^{(\ell)}\!(\tfrac1u,\tfrac vu),
 \qquad
 \Phi_v^{(\ell)}=\frac1v\Phi^{(\ell)}\!(\tfrac uv,\tfrac1v).
 \label{eq:appendix-crossed-ladders}
\end{equation}

In the source rest frame, let $T$ be the source separation, with $\operatorname{Re}T>0$, and $\tau_2,\tau_3$ the detector times, ordered as $\operatorname{Im}T<\operatorname{Im}\tau_2<\operatorname{Im}\tau_3<0$. Thus, the detector integrand~\eqref{eq:event-shape} can be written as
\begin{equation}
 \mathcal{I}_1^{(2)}(\tau_2,\tau_3)
 =-\frac{T^2}{64\pi^5\gamma}\,
 \frac{G_1^{(2)}(u,v)}
 {\tau_2(\tau_2-T)\tau_3(\tau_3-T)},
 \label{eq:appendix-k1-time-integrand}
\end{equation}
with
\begin{equation}
 u=-\frac{\gamma(\tau_2-T)\tau_3}{T^2},
 \qquad
 v=-\frac{\gamma(\tau_3-T)\tau_2}{T^2}.
 \label{eq:appendix-k1-time-arguments}
\end{equation}
where one integrates over the real parts of $\tau_2,\tau_3$.

For $\operatorname{Re}\tau_2=-R\xi$ and $\operatorname{Re}\tau_3=R(1-\xi)$, with $R\to\infty$ at fixed $0<\xi<1$, the Wightman prescription gives $\log v-\log u\to-2\pi i$. In this limit $u$ and $v$ grow like $R^2$, so the denominator in~\eqref{eq:two-loop-box-definition} is large except within $1-t\sim1/R$ of the endpoint, where it stays of order one. Hence $\Phi^{(1)}=O(1/R)$, and the leading term of the $\mathbf{84}$ channel~\eqref{eq:r-k1-channel-relation} is $\left((u+v)\Phi^{(1)}\right)^2$, which gives the integrand
\begin{equation}
 \begin{aligned}
 \mathcal{I}_1^{(2)}
 &=-\frac{1}{4\pi^3[\gamma-4\xi(1-\xi)]R^2}
 \log^2\!\frac{\sqrt{\gamma}+\sqrt{\gamma-4\xi(1-\xi)}}{\sqrt{\gamma}-\sqrt{\gamma-4\xi(1-\xi)}}
 \\
 &\quad+O\!\left(R^{-3}\log^4 R\right),
 \end{aligned}
 \label{eq:appendix-k1-time-tail}
\end{equation}
for $\xi$ away from $0$ and $1$. Since the measure supplies another $R\,dR\,d\xi$, this tail scales as $\int dR/R$, generating a logarithmic divergence. 

\section{Two-loop event shapes at large \texorpdfstring{$k$}{k}}
\label{ap:two-loop-detectors}
We work above saturation ($k\geq3$) and fixed angles $0<\theta<\pi$, where the two-loop correlator is 
\begin{equation}
  G_k^{(2)}=(\Phi^{(1)})^2+4\Phi^{(2)}.
\end{equation}

Writing $x=-z/(1-z)$ and $y=(\bar z-1)/\bar z$ maps the cuts to $0<x,y<1$, so the mixed discontinuity gives
\begin{equation}
 \operatorname{Disc}_z\operatorname{Disc}_{\bar z}G_k^{(2)}
 =4\pi^2\left[\frac{H(x,y)}{(z-\bar z)^2}
              +\frac{J(x,y)}{z-\bar z}\right],
 \label{eq:two-loop-mixed-discontinuity}
\end{equation}
where
\begin{align}
 H(x,y)&=\frac12\left[\log^2\frac{x(1-y)}{y(1-x)}
                    -\log^2x-\log^2y\right]\nonumber\\
 &\quad-\zeta(2)-\operatorname{Li}_2(x)-\operatorname{Li}_2(y),
 \nonumber\\
 J(x,y)&=2\zeta(2)-\operatorname{Li}_2(x)-\operatorname{Li}_2(y)
                  +\log x\log y.
 \label{eq:two-loop-cut-functions}
\end{align}
The two terms in~\eqref{eq:two-loop-mixed-discontinuity} come from $(\Phi^{(1)})^2$ and $4\Phi^{(2)}$, respectively. Inserting them into~\eqref{eq:mixed-discontinuity-kernel} gives
\begin{align}
 \mathcal{G}_k^{(2)}(\gamma)
 &=\frac{4k\gamma^{2k-2}}{\pi^3 16^k}
 \int_0^1dx\int_0^1dy\,
 \frac{[(1-x)(1-y)]^{2k-2}}{(xy)^k}
 \nonumber\\
 &\quad\times\left[J(x,y)
 -\frac{(1-x)(1-y)}{1-xy}H(x,y)\right]
 \nonumber\\
 &\quad\times\mathcal{R}_{u,k}\mathcal{R}_{v,k}
 \left[\frac{1}{\sqrt{(\gamma-u-v)^2-4uv}}\right],
 \label{eq:two-loop-finite-charge-integral}
\end{align}
where one replaces the cross-ratios $u=-x/[(1-x)(1-y)]$ and $v=-y/[(1-x)(1-y)]$ after applying the Taylor remainder operators $\mathcal{R}_{x,k}$.

These differential operators have a uniform large-charge expansion for $u,v<0$
\begin{multline}
 \mathcal{R}_{u,k}\mathcal{R}_{v,k}
 \left[\frac{1}{\sqrt{(\gamma-u-v)^2-4uv}}\right]
 \\
 =\frac{16^k\gamma}{\pi k(\gamma-4u)(\gamma-4v)}
 \left(\frac{uv}{\gamma^2}\right)^k\left[1+O(k^{-1})\right].
 \label{eq:two-loop-remainder-asymptotic}
\end{multline}
Substituting this expansion into~\eqref{eq:two-loop-finite-charge-integral} and integrating gives
\begin{multline}
 \mathcal{G}_k^{(2)}(\gamma)
 =\frac{4}{\pi^4\gamma}
 \int_0^1dx\int_0^1dy\,
 \\
 \times\frac{
 J(x,y)-\dfrac{(1-x)(1-y)}{1-xy}H(x,y)
 }{
 [\gamma(1-x)(1-y)+4x][\gamma(1-x)(1-y)+4y]
 }
 \\+O(k^{-1}).
 \label{eq:two-loop-large-charge-integral}
\end{multline}
and after integrating the $H$ and $J$ terms reduces to
\begin{multline}
 \mathcal{G}_k^{(2)}
 =\frac{1}{\pi^4\gamma}\Biggl(
 4\left[\int_0^1\frac{\log t\,dt}{\gamma(1-t)^2+4t}\right]^2
 \\
 -\int_0^1\frac{\log^3t\,dt}{\gamma(1-t)^2+4t}\Biggr)+O(k^{-1}).
 \label{eq:two-loop-logarithmic-integrals}
\end{multline}
Evaluating these integrals and taking $k\to\infty$ gives
\begin{equation}
 \lim_{k\to \infty} \mathcal{G}_k^{(2)}
 =\frac{3\sin^{4}(\theta/2)}{\pi^{4}\cos^{2}(\theta/2)}
 \left[\frac{\operatorname{Cl}_{2}(\theta)^{2}}{3}+\frac{\operatorname{Cl}_{4}(\theta)}{\tan(\theta/2)}\right],
 \label{eq:two-loop-heavy-local-evaluation}
\end{equation}
which reproduces~\eqref{eq:large-k-two-loop-event-shape}.

\section{The large-charge limit}
\label{ap:limits-interchange}
We now take the large-charge limit at fixed positive loop order $\ell$ and fixed angle $0<\theta\leq\pi$, with the contours $j_a=c+it_a$, $-1/2<c<0$, used in Appendix~\ref{ap:detector-transform}. The integrable bound~\eqref{eq:appendix-mellin-majorant} allows us to pass this limit through the kernel's Mellin integral. At fixed Mellin variables,
\begin{equation}
 K_k(j_1,j_2)
 =\frac{4^{-j_1-j_2}}{16\pi^4}\bigl[1+O(k^{-1})\bigr].
 \label{eq:appendix-fixed-mellin-limit}
\end{equation}
The remaining Mellin integrals factorize, and each follows from
\begin{equation}
 \int_{c-i\infty}^{c+i\infty}\frac{dj}{2\pi i}\,
 \frac{(\gamma/(-4u))^j}{\sin(\pi j)}
 =\frac{4u}{\pi(\gamma-4u)}.
 \label{eq:appendix-cauchy-mellin}
\end{equation}
Substitution recovers the rational kernel~\eqref{eq:rational-kernel}.

The bound~\eqref{eq:appendix-full-majorant} then permits taking the limit inside the detector integral. Dominated convergence gives the dispersion relation~\eqref{eq:heavy-dispersion} at order $\ell\geq1$,
\begin{equation}
 \lim_{k\to\infty}\mathcal{G}_k^{(\ell)}
 \!=\!-\tfrac{1}{\pi^6\gamma}\!\iint\limits_{(-\infty,0)^2}\!du\,dv\,
 \frac{\operatorname{Disc}_z\!\operatorname{Disc}_{\bar z}
 [\mathbb{O}^2]^{(\ell)}}{(\gamma-4u)(\gamma-4v)}.
 \label{eq:appendix-dispersion}
\end{equation}
To evaluate~\eqref{eq:appendix-dispersion}, we restrict to $-1/4<c<0$ and justify exchanging the correlator's Mellin integral with the two position integrations. Let $M^{(\ell)}$ denote the Mellin amplitude of the saturated correlator $[\mathbb{O}^2]^{(\ell)}$. In Euclidean kinematics, the ladder representations of~\cite{Kniehl:2013dma}, after Mellin integration by parts to absorb powers of $\log(u/v)$, express each ladder's Mellin amplitude as a finite sum of derivatives of the box Gamma factors. On contours $j_a=d+it_a$ with $-1/4<d<0$, Stirling's formula gives
\[
 \begin{aligned}
 &e^{\pi(|t_1|+|t_2|)}
 \left|\Gamma(-j_1)^2\Gamma(-j_2)^2
 \Gamma(1+j_1+j_2)^2\right|
 \\
 &\qquad\leq C_d\,
 \frac{e^{-\pi|t_1+t_2|}(1+|t_1+t_2|)^{2+6d}}
 {(1+|t_1|)^{2+4d}}.
 \end{aligned}
\]
Integrating this bound over $t_2$ leaves a constant times $(1+|t_1|)^{-2-4d}$, which is integrable since $d>-1/4$. Derivatives of the Gamma factors add at most powers of logarithms and preserve this convergence. The weighted integral is therefore finite for every ladder.

For a product of $m$ ladders, the Mellin variables add. We therefore choose real part $d=c/m$ for each factor, so that their sums have real part $c$. The product's Mellin amplitude is the convolution of the individual amplitudes. Writing the product's Mellin variables as $j_a=c+it_a$, with $a=1,2$, each $t_a$ is the sum of the corresponding imaginary parts of the factors. The triangle inequality therefore bounds the weight $e^{\pi(|t_1|+|t_2|)}$ by the product of the single-ladder weights. Taking absolute values and integrating then gives a bound by the product of the corresponding single-ladder integrals. These are finite by the preceding estimate, since $-1/4<c/m<0$. At fixed $\ell$, the saturated correlator is a finite sum of such products, so
\begin{equation}
 \int_{\mathbb{R}^2}\!dt_1\,dt_2\,
 e^{\pi(|t_1|+|t_2|)}
 \left|M^{(\ell)}(c+it_1,c+it_2)\right|<\infty.
 \label{eq:appendix-correlator-mellin-bound}
\end{equation}
The sine factors in~\eqref{eq:mixed-disc-mode} are bounded by this exponential weight, and the position integrals of the absolute Mellin powers converge for $-1<c<0$. Thus~\eqref{eq:appendix-correlator-mellin-bound} justifies exchanging the correlator's Mellin integral with both integrations in~\eqref{eq:appendix-dispersion}. We then integrate each Mellin power with the identity
\begin{equation}
 \int_{-\infty}^0du\,\frac{(-u)^j}{\gamma-4u}
 =-\frac{\pi(\gamma/4)^j}{4\sin(\pi j)},
 \label{eq:appendix-cauchy-power}
\end{equation}
valid for $-1<\operatorname{Re}j<0$.
Applying~\eqref{eq:appendix-cauchy-power} twice cancels the sine factors and evaluates the correlator at $u=v=\gamma/4$. Under~\eqref{eq:octagon-limit}, this point maps to $z_{\text{o}}=e^{-i\theta}$ and $\bar z_{\text{o}}=e^{i\theta}$.
Together with~\eqref{eq:heavy-tree-level-event-shape}, this proves~\eqref{eq:mainResult1} and~\eqref{eq:mainResult} order by order in perturbation theory.

\end{document}